\documentclass[12pt]{article}
\usepackage{amsfonts}
\usepackage{amsmath}
\usepackage{amssymb}
\usepackage{graphicx}
\usepackage{natbib}
\usepackage{hyperref}
\usepackage{authblk}

\usepackage{graphicx}
\usepackage{color}%
\usepackage[utf8]{inputenc}

\begin{document}

\title{Robust Estimation in High Dimensional Generalized Linear Models}
\author[1]{M. Valdora\thanks{Corresponding author mvaldora@gmail.com}}
\author[2]{C. Agostinelli}
\author[3]{V.J. Yohai}
\affil[1]{Departamento de Matematicas, Facultad de Ciencias Exactas y Naturales, University of Buenos Aires}
\affil[2]{Department of Mathematics, University of Trento, Trento, Italy}
\affil[3]{Departamento de Matematicas, Facultad de Ciencias Exactas y Naturales, University of Buenos Aires and CONICET}

\date{\today}

\maketitle

\begin{abstract}
Generalized Linear Models are routinely used in data analysis. The classical procedures for estimation are based on Maximum Likelihood and it is well known that the presence of outliers can have a large impact on this estimator. Robust procedures are presented in the literature but they need a robust initial estimate in order to be computed. This is especially important for robust procedures with non convex loss function such as redescending M-estimators. Subsampling techniques are often used to determine a robust initial estimate; however when the number of unknown parameters is large the number of subsamples needed in order to have a high probability of having one subsample free of outliers become infeasible. Furthermore the subsampling procedure provides a non deterministic starting point. Based on ideas in \citet{PenaYohai1999}, we introduce a deterministic robust initial estimate for M-estimators based on transformations \citep{ValdoraYohai2014} for which we also develop an iteratively reweighted least squares algorithm. The new methods are studied by Monte Carlo experiments.

\end{abstract}

\section{Introduction}

We consider high dimensional generalized linear models and we study a robust
method for estimating its parameters. Robust estimators for generalized linear
models (GLM) have been studied by \citet{Kunschetal1989},
\citet{CantoniRonchetti2001}, \citet{BergesioYohai2011},
\citet{Biancoetal2013}, \citet{ValdoraYohai2014} and
\citet{AlqallafAgostinelli2016}. However, these proposals either lack
robustness or require a robust initial estimator. We propose a method for
computing an initial estimator which can be used to begin an iterative
algorithm, as needed by redescending estimators. We apply this method in the
computation of M-estimators based on transformations (MT) proposed by  \citet{ValdoraYohai2014}. MT-estimators are a family of M-estimators based on variance stabilizing transformations which are shown to be highly robust and efficient by means of a Monte Carlo study. These estimators are redescending M-estimators applied after transforming the responses by means of a variance stabilizing function. Stabilizing the variance allows the correct scaling of the loss function used in the definition of the M-estimator.

Consider a GLM in which $y$ is the response and $\mathbf{x}$ is a
$p$-dimensional vector of explanatory variables. We assume that
\begin{equation}
\label{link}
g(\mu)=\boldsymbol{\beta}_{0}^{\top}\mathbf{x},
\end{equation}
where $\boldsymbol{\beta}_{0}\in\mathbb{R}^{p}$ is an unknown vector of
parameters and $g:$ $\mathbb{R}\rightarrow$ $\mathbb{R}$ is a known link
function. We further assume that
\begin{equation}
\label{condmod}
y|\mathbf{x\sim}F_{\mu},
\end{equation}
where $F_{\mu}$ is a discrete or continuous model in the exponential family of
distributions in $\mathbb{R}$, that is to say, it has a density of the form
\begin{equation}
\label{eq:expfam}
f_{\mu}(y) =\exp(\left(  \left(  y\mu- b(\mu)\right) /a(\phi)+c(y,\phi)\right),
\end{equation}
for given functions $a$, $b$ and $c$. We assume $\phi$ is known. MT-estimators are defined as follows
\begin{align}
L(\boldsymbol{\beta)} & = \sum_{i=1}^{n} \rho\left( t(y_{i}) - m\left( g^{-1}\left( \mathbf{x}_{i}^{\top} \boldsymbol{\beta}) \right) \right) \right) \nonumber \\
\hat{\boldsymbol{\beta}} & = \arg\min_{\boldsymbol{\beta}} L(\hat{\boldsymbol{\beta}}) \label{Mest1}
\end{align}
where $\rho(u)$ is a symmetric, bounded, continuous and non decreasing on
$|t|$ function, $t$ is a variance stabilizing transformation and $m$ is the
function defined by
\begin{equation}
m(\mu)=\text{argmin}_{\gamma}\mathbb{E}_{\mu}\left( \rho\left(t(y)-\gamma\right) \right), \label{mmu}
\end{equation}
where $\mathbb{E}_{\mu}(y)$ denotes the expectation of $y$ when $y$ has distribution $F_{\mu}$.
It is assumed that $m$ is univocally defined, therefore (\ref{mmu}) implies the Fisher consistency of $\hat{\boldsymbol{\beta}}$.  Other assumptions
necessary to have consistency and asymptotic normality of this estimators are
listed in \citet{ValdoraYohai2014}. The solution to (\ref{Mest1}) can be
found by iterative methods which typically solve the corresponding system of
estimating equations
\begin{equation}
\sum_{i=1}^{n}\psi(\mathbf{x}_{i},y_{i},\boldsymbol{\beta})=0. \label{eq:Zest}
\end{equation}
where $\psi(\mathbf{x}_{i},y_{i},\boldsymbol{\beta})$ is the derivative with
respect to $\boldsymbol{\beta}$ of $\rho\left(  t(y_{i})-m\left(
g^{-1}\left(  \mathbf{x}_{i}^{\top}\boldsymbol{\beta})\right)  \right)
\right)$.
In 
the Appendix we provide an iteratively reweighted least squares (IRWLS) algorithm to find a solution to equation (\ref{eq:Zest}). The difficulty in the case of redescending M-estimators is that the
goal function $L(\boldsymbol{\beta})$ might have several local minima. As a
consequence, it might happen that the iterative procedure converges to a
solution of equation (\ref{eq:Zest}) that is not a solution of the optimization problem (\ref{Mest1}). To
avoid this, one must begin the iterative algorithm at an initial estimator
which is a very good approximation of the absolute minimum of $L$, i.e. the solution of (\ref{Mest1}).
If $p$ is small, this approximate solution may be obtained by the subsampling
method \citet[see]{ValdoraYohai2014}. Based on the algorithm described in
\citet{RousseeuwLeroy1987} for linear models, this method consists in
computing a finite set $A$ of candidate solutions to (\ref{Mest1}) and then
replace the minimization over $\mathbb{R}^{p}$ by a minimization over $A$. The
set $A$ is obtained by randomly drawing subsamples of size $p$ and computing
the maximum likelihood (ML) estimator based on the subsample. If the original
sample contains a proportion $\epsilon$ of outliers, then the probability that
a given subsample is free of outliers is $(1-\epsilon)^{p}$ and the
probability of having at least one subsample free of outliers is
$1-(1-(1-\epsilon)^{p})^{N}$, where $N$ is the number of subsamples drawn. If
we want this probability to be greater than a given $\alpha$, we must draw a
number of subsamples such that
\[
1-(1-(1-\epsilon)^{p})^{N}>\alpha,
\]
that is to say,
\[
N>\frac{\log(\alpha)}{\log(1-(1-\varepsilon)^{p})}\underline{\sim}\left\vert
\frac{\log(\alpha)}{(1-\varepsilon)^{p}}\right\vert .
\]
This makes the algorithm infeasible for large $p$.

\citet{PenaYohai1999} studied this problem in the case of linear models,
introducing an alternative method to compute the set of candidate solutions
$A$. Their proposal succeeds in obtaining a set $A$ which contains very good
approximations of the actual solution and, on the other hand, requires the
computation of a small number of subsamples, namely $3p+1$. This makes the
algorithm much faster and feasible even for very large values of $p$.

We modify the method introduced by \citet{PenaYohai1999} in order to apply it
to generalized linear models. 
We study its application to MT-estimators by
means of an extensive Monte-Carlo study, which shows that the method is very
fast and robust for large values of $p$.

As a particular case of the MT-estimator we define the Least Squares estimator
based on Transformations (LST), which corresponds to $\rho(u)=u^{2}$, in the following way
\begin{equation}
\hat{\boldsymbol{\beta}}=\text{argmin}_{\boldsymbol{\beta}}\sum_{i=1}%
^{n}\left(  t(y_{i})-\mathbb{E}_{\left(  g^{-1}\left(  \mathbf{x}_{i}^{\top
}\boldsymbol{\beta})\right)  \right)  }\left(  t(y_{i})\right)  \right)  ^{2}.
\label{eq:LSest}%
\end{equation}
This estimator can be seen as a natural generalization of the Least Squares estimator (LS) for linear models to the case of GLM. LST estimators are Fisher consistent, however since $\rho$ is not bounded, they are, in general, non robust. In 
the Appendix we provide an iteratively reweighted least squares algorithm to find the solution to the optimization problem (\ref{eq:LSest}).

\section{Detecting Outliers Using Principal Sensitivity Components}

The classical statistic used to measure the influence of an observation is the
Cook statistic introduced by \citet{Cook1977} for linear models, which can be
adapted for generalized linear models (see Chapter 12 of
\citet{McCullaghNelder1989}). This statistic is a measure of the distance
between $\hat{\boldsymbol{\beta}}$, the maximum likelihood estimator and
$\hat{\boldsymbol{\beta}}_{(i)}$, the maximum likelihood estimator computed
without observation $i$. However, these measure is non-robust and therefore,
when there are several ouliers, it may be completely unreliable. In these
cases, some outliers $(y_{i},\mathbf{x}_{i})$ with high influence may have a
small Cook statistic if there are other similar outliers that still influence
$\hat{\boldsymbol{\beta}}_{(i)}$. This is known as masking effect. To make
things worse, high leverage outliers may have small residuals making their detection difficult. This
situation usually arises when there are several similar, or highly correlated outliers.

The proposal of Pe\~na and Yohai (1999) follows the same idea as the
subsampling method but it computes the set of candidate solutions $A$ in a
different way. The candidates are obtained, as before, by computing the least
squares estimates on subsamples. However, the subsamples are not chosen at
random. Instead, they are chosen by deleting from the sample, groups of similar
or highly correlated outliers, which can potentially cause a masking effect.
The set $A$ will, in this way, contain candidates which are already quite
robust estimates and therefore it will not need to have a large number of
candidates as it happens using randomly chosen subsamples. In fact the number of candidates in the set $A$ is only $3p+1$.

Let $(\mathbf{x}_{1},y_{1}),\dots,(\mathbf{x}_{n},y_{n})$ be random vectors
which follow a generalized linear model as defined by (\ref{condmod}) and
(\ref{link}). Let $\hat{\boldsymbol{\beta}}$ be the LST estimator and
\[
\hat{\boldsymbol{\mu}}=(\hat{\mu}_{1},\ldots,\hat{\mu}_{n})^{\top}=g^{-1}%
(\mathbf{X}\hat{\boldsymbol{\beta}})
\]
be the vector of fitted values. Let $\hat{\mu}_{i(j)}$ be the fitted value for
observation $i$ computed without using observation $j$, that is $\hat{\mu
}_{i(j)}=g^{-1}(\mathbf{x}_{i}^{\top}\boldsymbol{\hat{\beta}}_{(j)})$, where
$\boldsymbol{\hat{\beta}}_{(j)}$ is the LST estimate based on the original
sample without observation $j$. We define the $i$-th residual $e_{i} $ as the
difference between $t_{i}=t(y_{i})$ and its predicted value $\hat{t}%
_{i}=m(g^{-1}(\mathbf{x}_{i}^{\top}\widehat{\boldsymbol{\beta}}))$,
that is $e_{i}=t_{i}-\hat{t}_{i}$. Following the ideas introduced by Pe\~{n}a
and Yohai (1999) for linear models, we define the sensitivity vectors as the
vectors $\mathbf{r}_{i}$ with entries
\[
r_{ij}=\hat{t}_{i}-\hat{t}_{i(j)}%
\]
where $\hat{t}_{i(j)}=m(\hat{\mu}_{i(j)})$ is the predicted value of $t_{i}$ computed without using observation $j$.
Then, $r_{ij}$ is the sensitivity of the forecast of the $t_{i}$ to the
deletion of observation $j$ and the sensitivity vectors are defined by
\[
\mathbf{r}_{i}=\left( r_{i1}, \ldots, r_{in} \right) \ , \qquad 1 \leq i \leq n.
\]
The sensitivity matrix $\mathbf{R}$ is defined as the matrix whose rows are
the vectors $\mathbf{r}_{i}$. Let
\begin{equation}
\mathbf{v}_{1}=\mbox{argmax}_{||\mathbf{v}||=1}\sum_{i=1}^{n}\left(
\mathbf{v}^{\top}\mathbf{r}_{i}\right)  ^{2}. \label{eq1}%
\end{equation}
$\mathbf{v}_{1}$ is the direction in which the projections of the sensitivity
vectors is largest. Let
\begin{equation}
\mathbf{z}_{1}=\mathbf{R}\mathbf{v}_{1}; \label{eq:proj1}%
\end{equation}
then $\mathbf{z}_{1}$ is the vector whose entries are the terms of the sum in
(\ref{eq1}). Therefore, the largest entries in $\mathbf{z}_{1}$ correspond to
the largest terms in the sum in (\ref{eq1}), which in turn correspond to the
observations that have the largest projected sensitivity in the direction
$\mathbf{v}_{1}$.

In the same way, we can define recursively $\mathbf{v}_{i},$ $2\leq i\leq n$
as the solution of
\begin{align}
\mathbf{v}_{i}  &  =\mbox{argmax}_{||\mathbf{v}||=1}\sum_{i=1}^{n}\left(
\mathbf{v}^{\top}\mathbf{r}_{i}\right)  ^{2}.\label{eq2}\\
\text{ subject to }\mathbf{v}_{i}\mathbf{v}_{j}  &  =0\text{ for all }1\leq
j<i
\end{align}
The vectors $\mathbf{v}_{1}\dots\mathbf{v}_{p}$ are the directions in which
the projected sensitivity of the observations are the largest. The corresponding
projections
\begin{equation}
\mathbf{z}_{i}=\mathbf{R}\mathbf{v}_{i} \label{eq:proj2}%
\end{equation}
are called the principal sensitivity components. The entries of $\mathbf{z}%
_{i}$ are the projections of the sensitivity vectors on the direction
$\mathbf{v}_{i}$. Large entries correspond to observations whose projected
sensitivity in the direction $\mathbf{v}_{i}$ is large. Therefore, large
entries are considered potential outliers.

High leverage observations typically have large sensitivity because a small
change in the estimated slopes will cause a large change in the fitted values.
\citet{PenaYohai1999} prove that, in the case of linear models, if the sample
is contaminated with less than $(n-p+1)/(2n-p+1)$ high leverage outliers,
then, at least for one eigenvector, the coordinates corresponding to the
outliers have absolute value larger than the median.

\section{Procedure for obtaining a robust initial estimate in generalized linear models}

Consider a random sample following a generalized linear model as defined by
(\ref{link}), (\ref{condmod}) and (\ref{eq:expfam}). The
following procedure computes an approximation of $\boldsymbol{\beta}_{0}$ which will be used as an initial estimator in the IRWLS algorithm for the estimating equation (\ref{eq:Zest}). The procedure has two stages. Stage 1 aims at finding a highly
robust but possibly inefficient estimate and stage 2 aims at increasing its efficiency.

\noindent{\itshape Stage 1.} In this stage, the idea is to find a robust, but
possibly inefficient, estimate of $\boldsymbol{\beta}$ $\ $by an iterative
procedure. In each iteration $k\geq1$ we get
\begin{equation}
\label{eq:optLikAk}
\hat{\boldsymbol{\beta}}^{(k)}=\arg\min_{\boldsymbol{\beta}\in A_{k}%
}L(\boldsymbol{\beta}).
\end{equation}
In the first iteration ($k=1$) the set $A_{1}$ is constructed as follows. 
We begin by computing the LST estimate with the complete sample and the principal sensitivity components.
For
each principal sensitivity component $\mathbf{z}_{i}$ we compute three estimates by the
LST method. The first eliminating the half of the observations
corresponding to the smallest entries in $\mathbf{z}_{i}$, the second
eliminating the half corresponding to the largest entries in $\mathbf{z}_{i}$
and the third eliminating the half corresponding to the largest absolute
values. To these $3p$ initial candidates we add the LST estimate computed
using the complete sample, obtaining a set of $3p+1$ elements. Once we have
$A_{1}$ we obtain $\hat{\boldsymbol{\beta}}^{(1)}$ by minimizing
$L(\boldsymbol{\beta})$ over the elements of $A_{1}$.

Suppose now that we are on stage $k.$ Let $0<\alpha<0.5$ be a trimming
constant; in all our applications we set $\alpha=0.05$. Then, for $k>1$, we
first delete the observations $(i=1,\cdots,n)$ such that $y_{i}>F_{\hat
{\boldsymbol{\mu}}_{i}}^{-1}(1-\alpha/2)$ or $y_{i}<F_{\hat{\boldsymbol{\mu}%
}_{i}}^{-1}(\alpha/2)$ where $\hat{\boldsymbol{\mu}}_{i}=g^{-1}\left(
\mathbf{x}_{i}^{\top}\hat{\boldsymbol{\beta}}^{(k-1)}\right)  $ and, with the
remaining observations, we re-compute the LST estimator $\hat
{\boldsymbol{\beta}}_{\ \text{LTS}}^{(k)}$ and the principal sensitivity
components. Let us remark that, for the computation of $\hat{\boldsymbol{\beta
}}_{\ \text{LTS}}^{(k)}$ we have deleted the observations that have large residuals,
since $\hat{\boldsymbol{\mu}}_{i}$ is the fitted value obtained using $\hat{\boldsymbol{\beta}}^{(k-1)}$. In this way, while candidates
on the first step of the iteration are protected from high leverage outliers,
candidate $\hat{\boldsymbol{\beta}}_{\text{LTS} }^{(k)}$ is protected from low leverage
outliers, which may not be extreme entries of the $\mathbf{z}_{i}$.

Now the set $A_{k}$ will contain $\hat{\boldsymbol{\beta}}_{\ \text{LST}%
}^{(k)}$, $\hat{\boldsymbol{\beta}}^{(k-1)}$ and the $3p$ LST
estimates computed deleting extreme values according to the principal
sensitivity components as in the first iteration. $\hat{\boldsymbol{\beta}}^{(k)}$ is the element of $A_k$ minimizing $L(\boldsymbol{\beta})$.

The iterations will continue until $\hat{\boldsymbol{\beta}}^{(k)} \approx \hat{\boldsymbol{\beta}}^{(k-1)}$. Let $\hat{\boldsymbol{\beta}}_{1}$ be the final estimate obtained at this stage.

\noindent{\itshape Stage 2.} We first delete the observations $y_{i}$
($i=1,\cdots,n$) such that $y_{i}>F_{\hat{\boldsymbol{\mu}}_{i}}^{-1}%
(1-\alpha/2)$ or $y_{i}<F_{\hat{\boldsymbol{\mu}}_{i}}^{-1}(\alpha/2)$, where
$\hat{\boldsymbol{\mu}}_{i}=g^{-1}\left(  \mathbf{x}_{i}^{\top}\hat{\boldsymbol{\beta}}_1\right)$ and compute the LST estimate
$\hat{\boldsymbol{\beta}}^{(\ast)}$ with the reduced sample. Then for each of the
deleted observations we check whether $y_{i}>F_{\hat{\boldsymbol{\mu}}_{i}%
}^{-1}(1-\alpha/2)$ or $y_{i}<F_{\hat{\boldsymbol{\mu}}_{i}}^{-1}(\alpha/2)$,
where $\hat{\boldsymbol{\mu}}_{i}=g^{-1}\left(  \mathbf{x}_{i}^{\top}%
\hat{\boldsymbol{\beta}}^{(\ast)}\right)$. Observations which are not within
these bounds are finally eliminated and those which are, are restored to the
sample. With the resulting set of observations we compute the LST estimate $\hat
{\boldsymbol{\beta}}_{2}$ which is our proposal as a starting value for
solving the estimating equations of the MT-estimates.

\section{Monte Carlo Study}

In this section we report the results of a Monte Carlo study in which we compare the MT-estimator computed with the proposed initial estimate (FMT), to the robust quasi likelihood
estimator (RQL) proposed by \citet{CantoniRonchetti2001}, the Conditionally Unbiased Bounded Influence (CUBIF) estimator proposed by \citet{Kunschetal1989}, and the MT-estimate beginning at an intitial estimator computing by subsampling (SMT). 
For computing the RQL estimator, we used function \texttt{glmrob} from the R package \texttt{robustbase} with method "Mqle" and argument \texttt{weights.on.x} set to "robCov" so that weights based on robust Mahalanobis distance of the design matrix (intercept excluded) are used to downweight potential outliers in x-space. The Conditionally Unbiased Bounded Influence (CUBIF) estimators proposed by \citet{Kunschetal1989} was computed using an implementation kindly provided by Prof. Alfio Marazzi (personal communication); the implementation available in function \texttt{glmrob} from the R package \texttt{robustbase} with method "cubif" perform substantially bad and it is not reported here. For the computation of the SMT estimator, the number of subsamples was set to $2500$. Both FMT and SMT are computed using an iteratively reweighted least squares method described in the appendix. They only differ in the starting point. We study the case of Poisson regression and $\log$ link.

Let $\mathbf{x}=(1,\mathbf{x}^{\ast})$ be a random vector in $\mathbb{R}^{p}$
such that $\mathbf{x}^{\ast}$ has distribution $\mathcal{N}_{p-1}(\mathbf{0},\mathbf{I})$\ and let $y$ be a random variable such that
$y|\mathbf{x}\sim\mathcal{P}\left(  \exp(\boldsymbol{\beta}_{0}^{\top}\mathbf{x})\right)$. We consider $p=100$ and three different models.
In model 1 data are generated with $\boldsymbol{\beta}_{0}=\mathbf{e}_{2}$, in
model 2 $\boldsymbol{\beta}_{0}=2\mathbf{e}_{1}+\mathbf{e}_{2}$ and in model 3
$\boldsymbol{\beta}_{0}=2\mathbf{e}_{1}+1.5\mathbf{e}_{2}$, where
$\mathbf{e}_{i}$ is the vector of $\mathbb{R}^{p}$ with all entries equal to
zero except for the $i$-th entry which is equal to one. For each of these
models we simulate the case in which the samples do not contain outliers and
the case in which the samples have  $10$ per cent of identical outliers at point
$(\mathbf{x}_{0},y_{0})$. The outliers are located at $\mathbf{x}_{0}%
=\mathbf{e}_{1}+3\mathbf{e}_{2}$. The values of $y_{0}$ are taken in a grid ranging from
$\boldsymbol{\mu}_{0}-K_{1}$ to $\boldsymbol{\mu}_{0}+K_{2}$ where
$\boldsymbol{\mu}_{0}=\exp(\boldsymbol{\beta}_{0}^{\top}\mathbf{x})=\mathbb{E}_{\boldsymbol{\beta}_{0}}\left(  y|\mathbf{x}=\mathbf{x}_{0}\right)$. The values
$K_{1}$ and $K_{2}$ and the grid step are chosen so that the maximum mean squared error of our proposed estimator can be identified. For model 1 we also
consider high leverage outliers with $\mathbf{x}_{0,2}=\mathbf{e}_{1}+3\mathbf{e}_{2}+4\mathbf{e}_{3}$. We have considered samples of size $n=400$ an $n=1000$, but since the behaviour is similar we report the results only for the larger sample size.
Given an estimator $\hat{\boldsymbol{\beta}}$, we denote by MSE, the mean
squared error defined by $\mathbb{E}_{\boldsymbol{\beta}_{0}}(||\hat
{\boldsymbol{\beta}}-\boldsymbol{\beta}_{0}||^{2})$, where $||\cdot||$ denotes
the $L_{2}$ norm. We estimate the MSE by
\[
\widehat{\operatorname{MSE}}=\frac{1}{N}\sum_{j=1}^{N}||\hat{\boldsymbol{\beta
}}_{j}-\boldsymbol{\beta}_{0}||^{2},
\]
where $\hat{\boldsymbol{\beta}}_{j}$ is the value of the estimator at the
$j$-th replication and $N$ is the number of replications which was chosen
equal to $1000$.

\begin{figure}
\begin{center}
\includegraphics[scale=0.4]{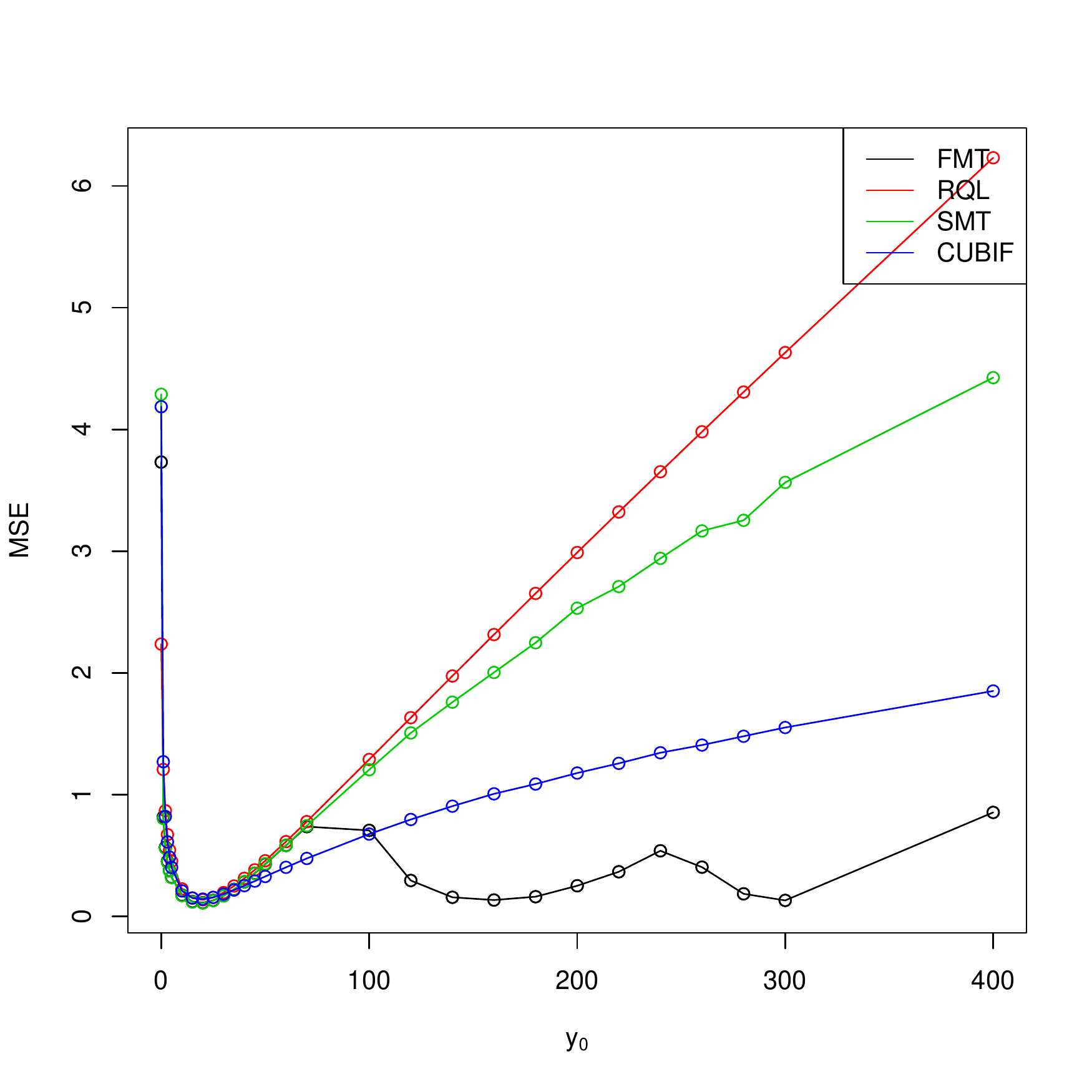}
\end{center}
\caption{MSE for model 1, $p=100$, $n=1000$ with $10\%$ outliers at $\mathbf{x}_0=3\mathbf{e}_{1}+\mathbf{e}_{2}$.}
\label{fig:model1}
\end{figure}

\begin{figure}
\begin{center}
\includegraphics[scale=0.4]{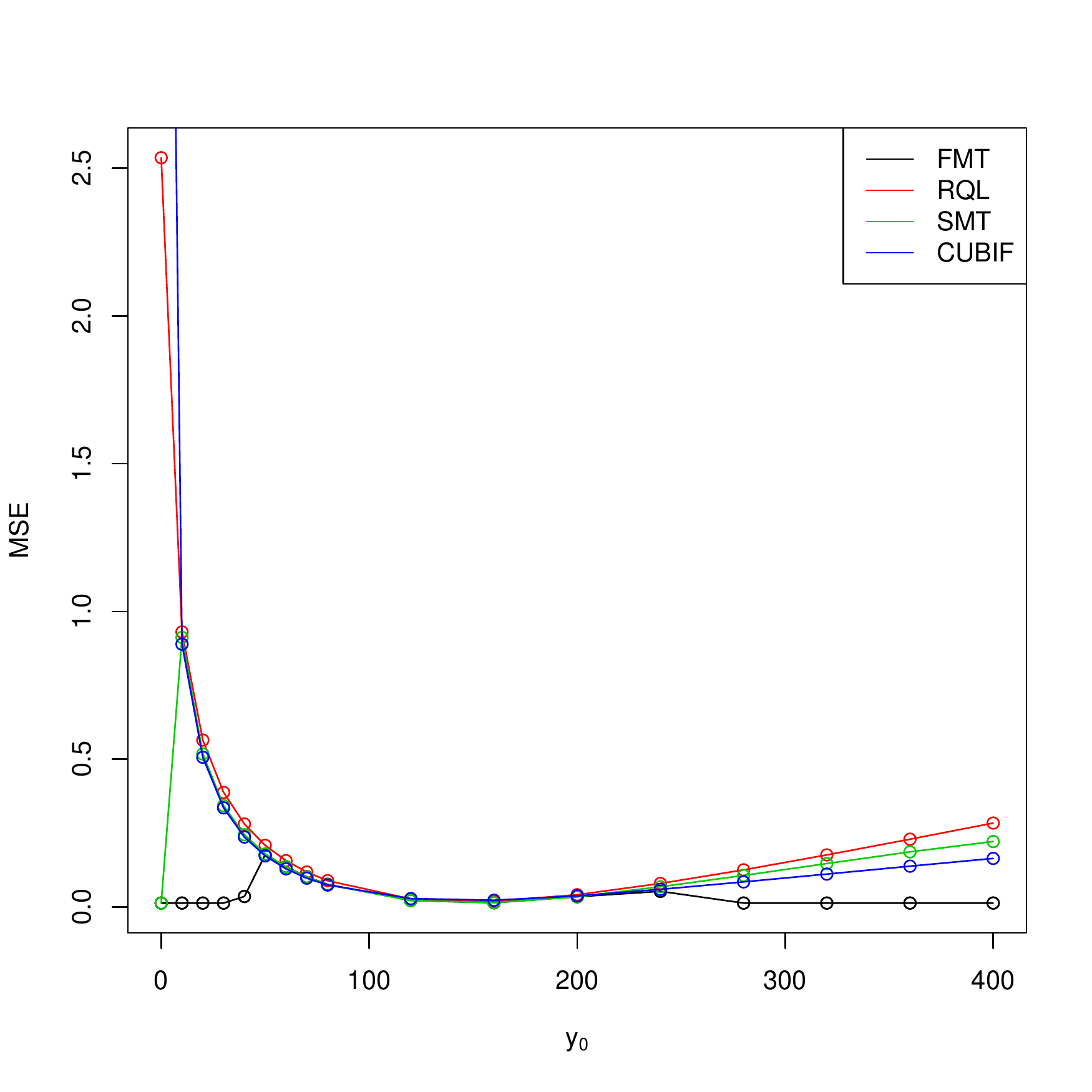}
\end{center}
\caption{MSE for model 2, $p=100$, $n=1000$ with $10\%$ outliers at $\mathbf{x}_0=3\mathbf{e}_{1}+\mathbf{e}_{2}$.}
\label{fig:model2}
\end{figure}

\begin{figure}
\begin{center}
\includegraphics[scale=0.4]{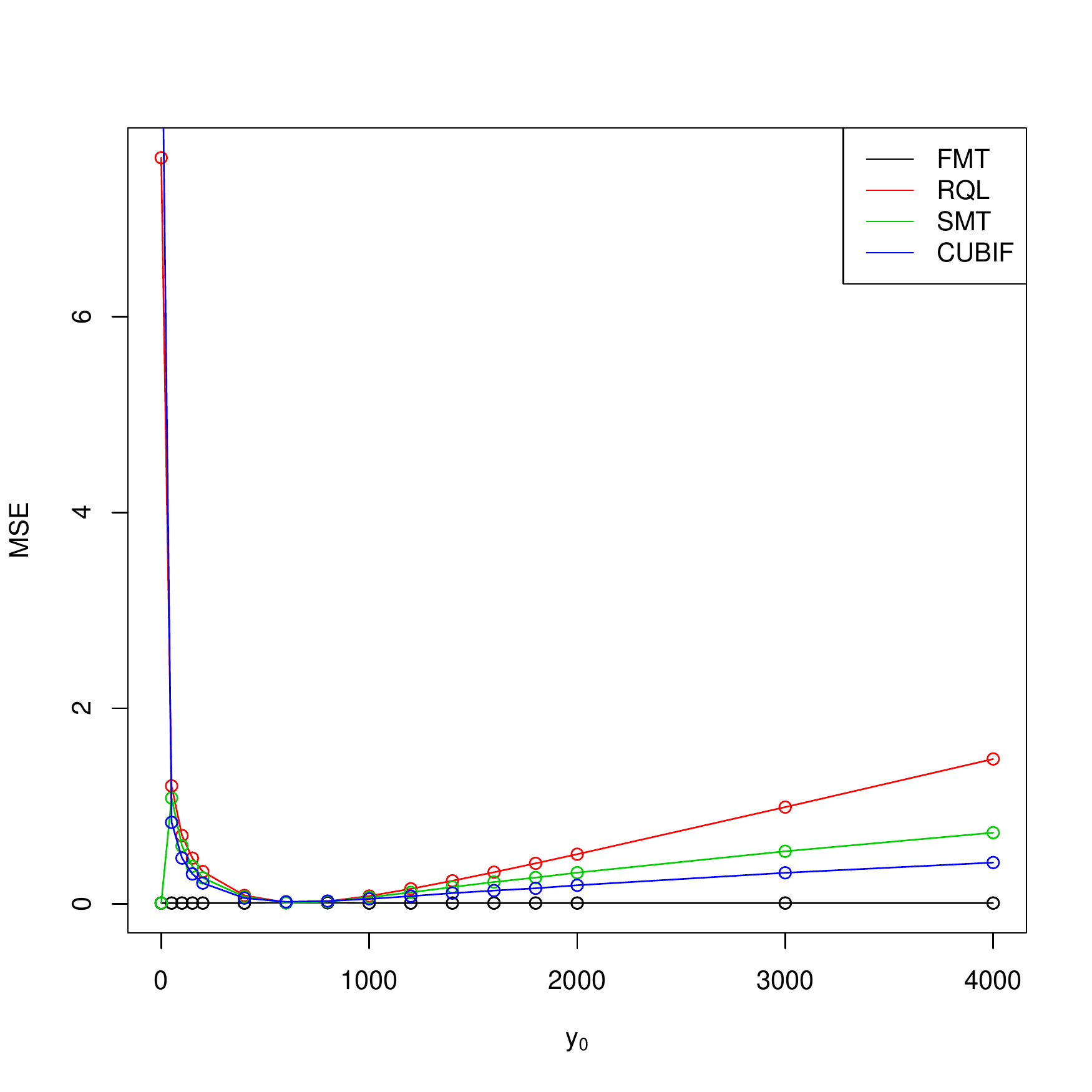}
\end{center}
\caption{MSE for model 3, $p=100$, $n=1000$ with $10\%$ outliers at $\mathbf{x}_0=3\mathbf{e}_{1}+\mathbf{e}_{2}$.}
\label{fig:model3}
\end{figure}

Our simulations show that the proposed estimator has smaller MSE than all
other proposals for almost all the contaminations considered. CUBIF estimator
has a smaller MSE for some values of $y_{0}$ but, since it is based on a
monotone score function, its MSE increases as $y_{0}$ increases. On the other
hand, the MSE of FMT estimator is bounded; we observe that it decreases
as $y_{0}$ increases beyond a certain value. To see this we study the MSE as a
function of $y_{0}$ and consider, as a measure of robustness, the maximum MSE
for $y_{0}\in\mathbb{Z}_{\geq0}$. The proposed estimator has the smallest
maximum MSE for all the models considered. 
In Figures \ref{fig:model1} to \ref{fig:model3} we plot the MSE as a function of $y_{0}$ for samples of size
$n=1000$ with $10\%$ contamination level. 

\begin{figure}
\begin{center}
\includegraphics[scale=0.3]{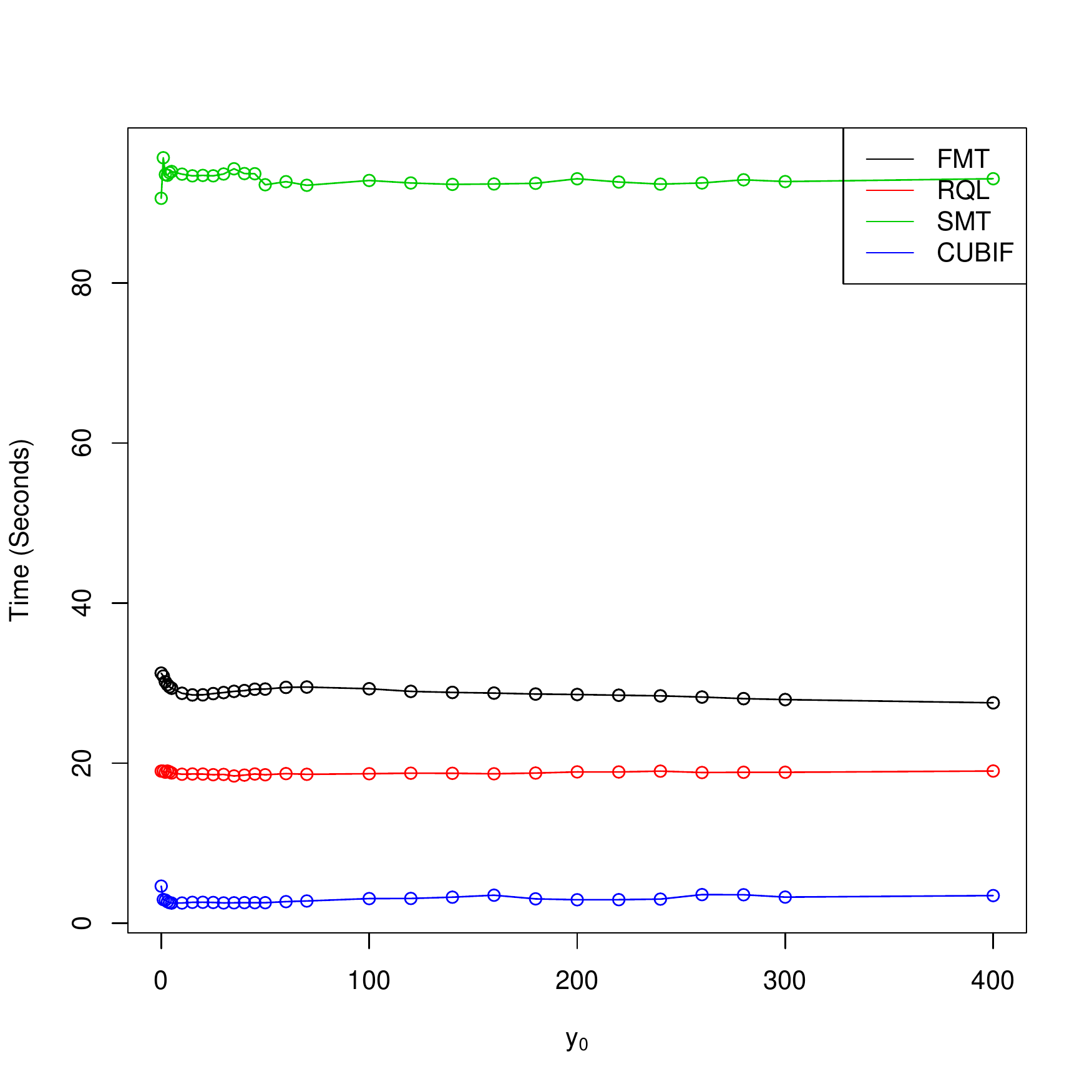}
\includegraphics[scale=0.3]{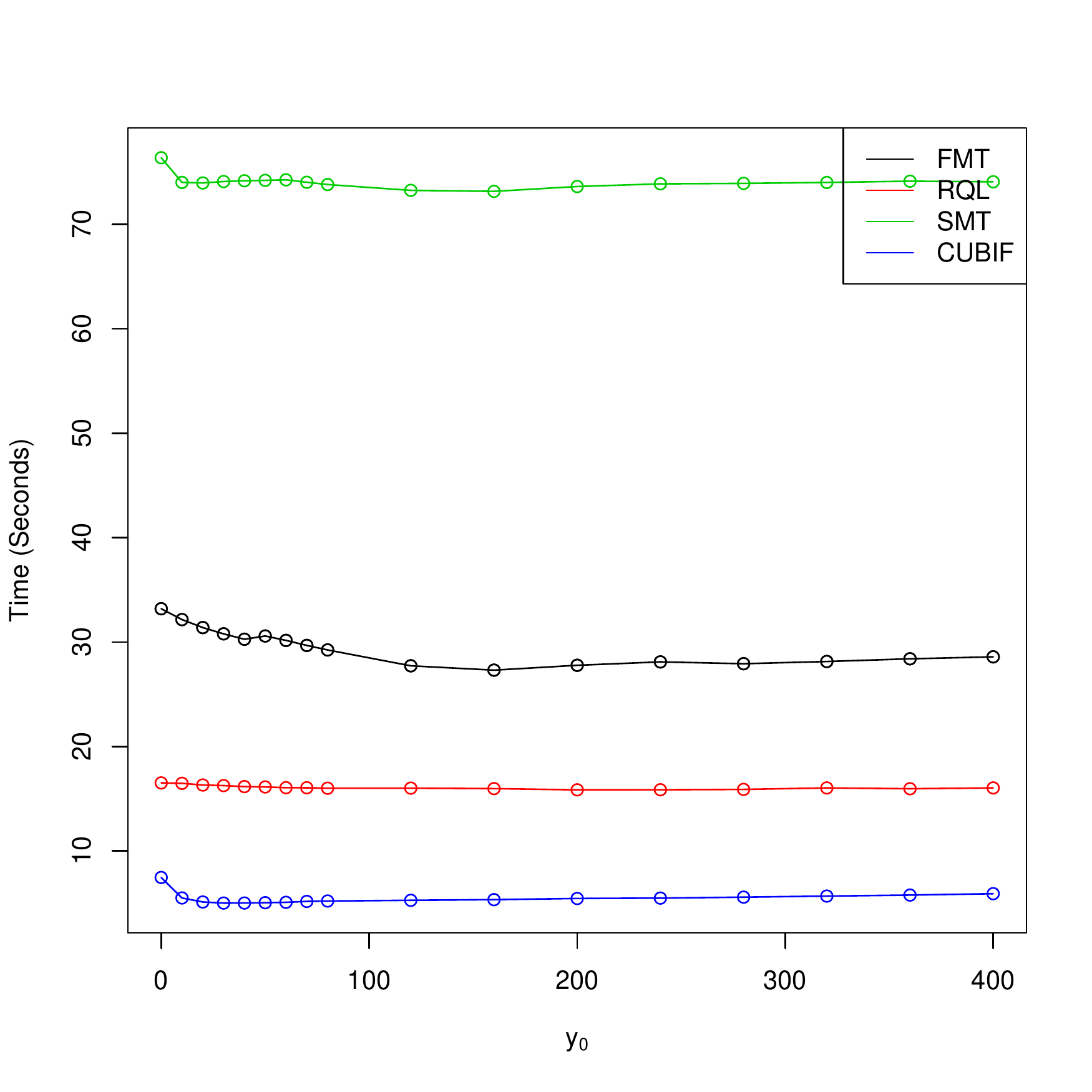}
\includegraphics[scale=0.3]{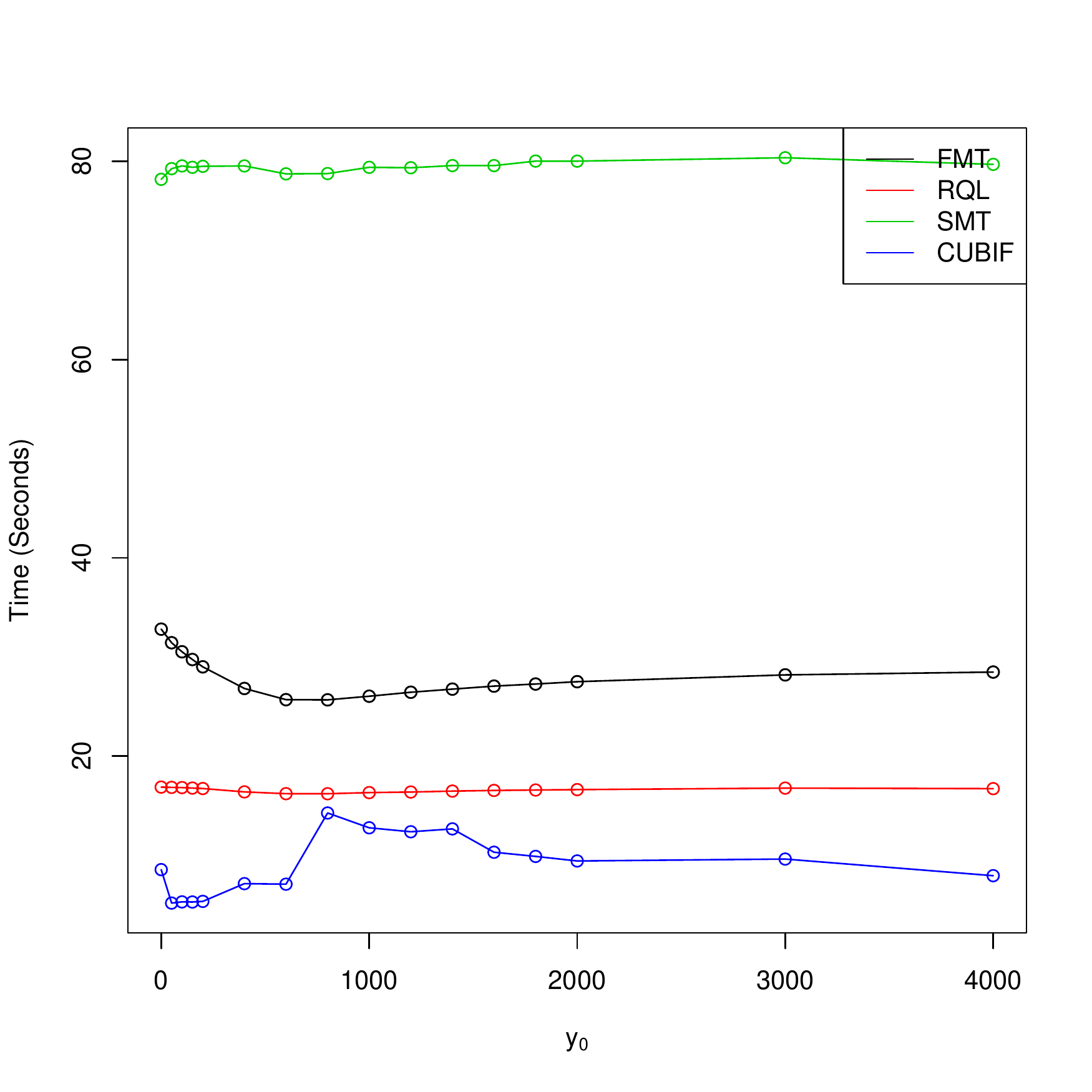}
\includegraphics[scale=0.3]{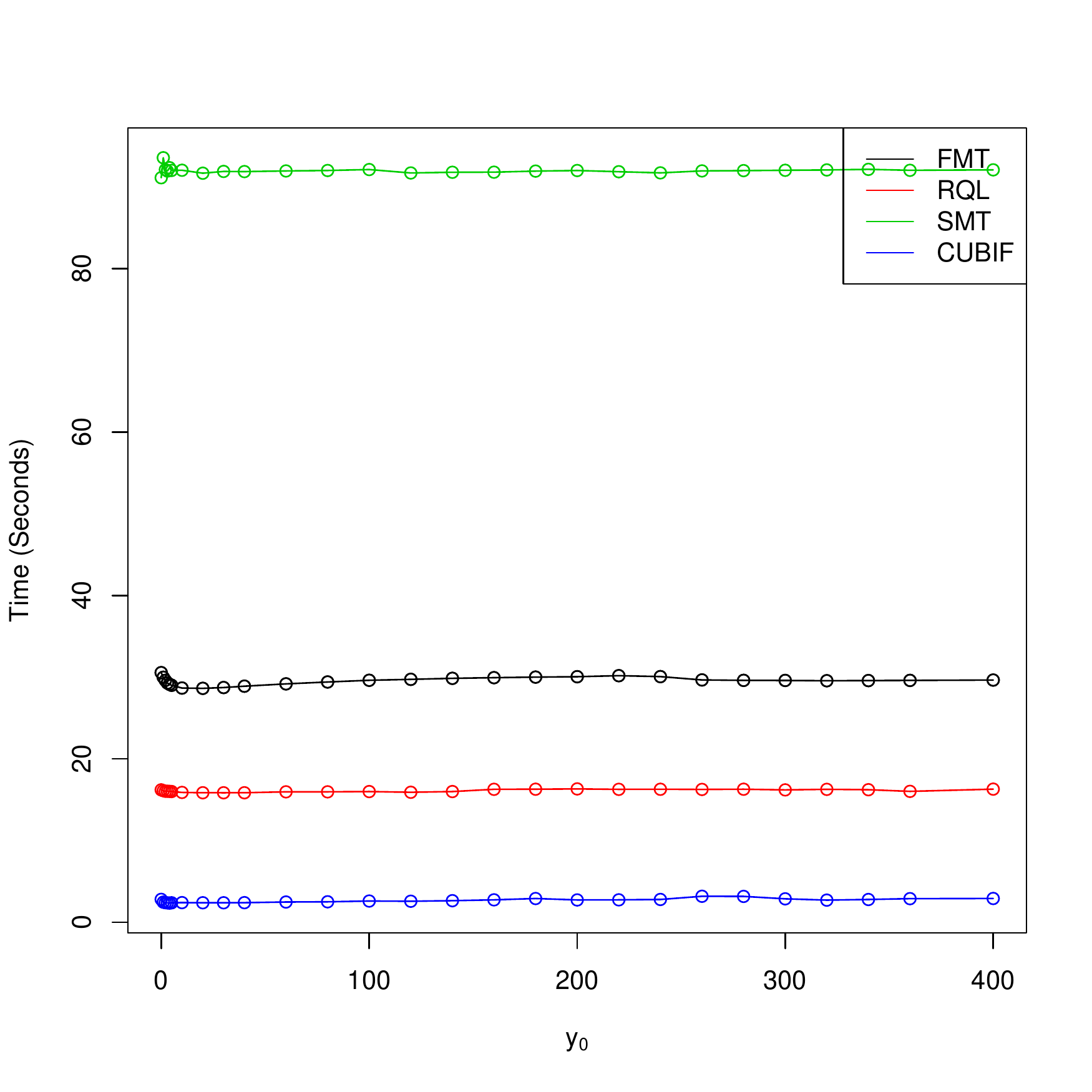}
\end{center}
\caption{Execution time in seconds for the four models, first row model 1 and 2, second row model 3 and 4, $p=100$, $n=1000$ with $10\%$ outliers.}
\label{fig:time}
\end{figure}

In Figure \ref{fig:time} we report the execution time for the different methods.
This figure shows that our proposed method is a great improvement over the subsampling method, as far as computational time is concerned. 

\begin{figure}
\begin{center}
\includegraphics[scale=0.4]{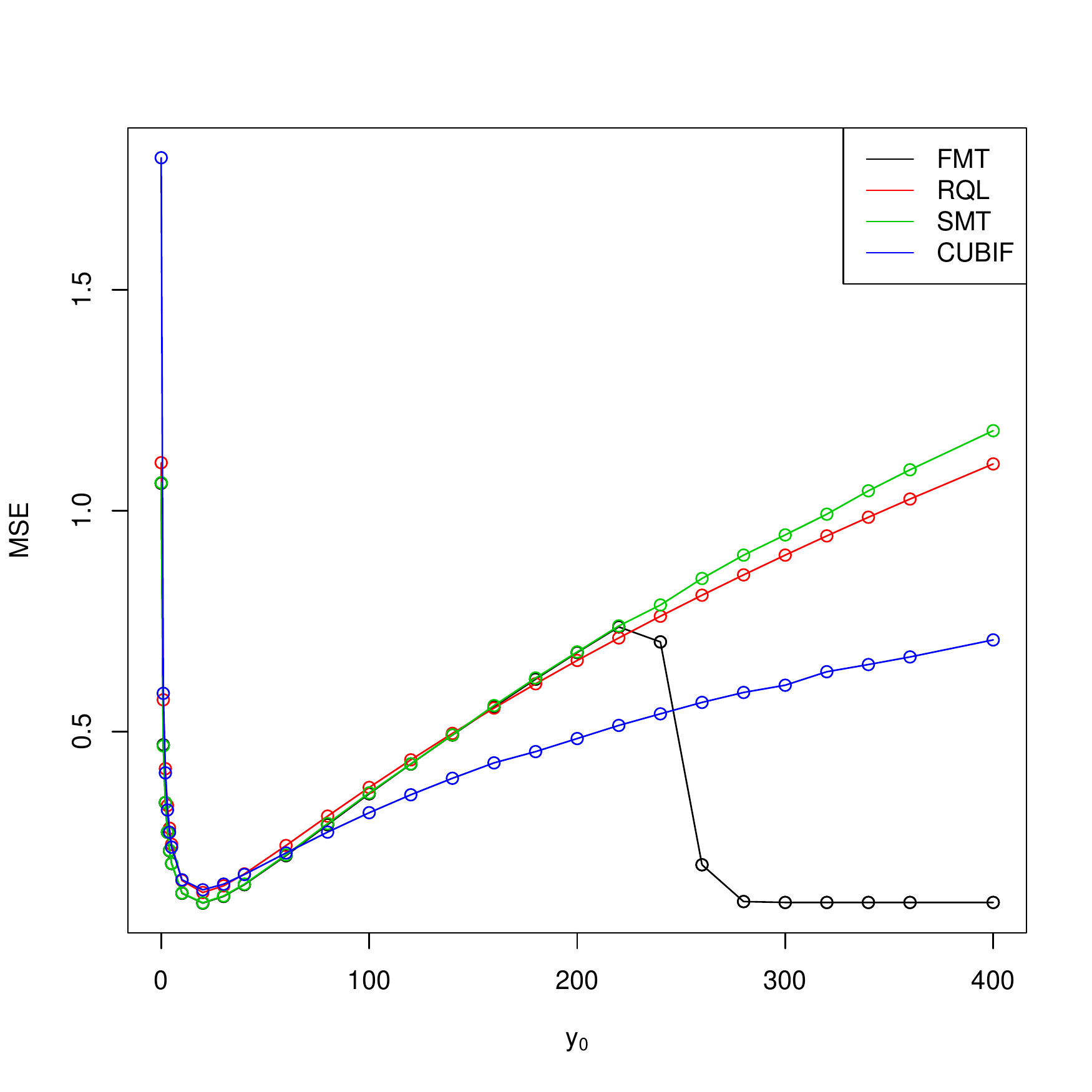}
\end{center}
\caption{MSE for model 4, $p=100$, $n=1000$ with $10\%$ outliers at $\mathbf{x}_0=3\mathbf{e}_{1}+\mathbf{e}_{2}+4\mathbf{e}_{4}$.}
\label{fig:model4}
\end{figure}

\section{Conclusion}
We introduce a deterministic robust initial estimate for generalized linear models. This initial estimate is used in an iteratively reweighted least squares algorithm to obtain a solution of a transformed M-estimator. We illustrate the procedures for the Poisson model. Monte Carlo experiments show that the performance of MT-estimators computed with the proposed initial estimator have a small bounded mean squared error exhibiting a redescending behavior. This is not the case for other proposals such RQL, CUBIF and MT with initial estimator based on subsampling.

\appendix

\section*{Computational details and algorithms}
\label{sec:computational}
In this Appendix we describe the iteratively reweighted least squares algorithms
that were used to compute the LST and the MT estimators.

Suppose that we have an initial estimator $\boldsymbol{\beta}_{0}$ and call
$s(t)=m(g^{-1}(t)),$ then using a Taylor expansion of order one we can
approximate $\ m\left(  g^{-1}\left(  \mathbf{x}_{i}^{\top}\boldsymbol{\beta
}\right)  \right)  =$ $s\left(  \mathbf{x}_{i}^{\top}\boldsymbol{\beta
}\right)  $ by%
\begin{equation}
\ \boldsymbol{s(}\mathbf{x}_{i}^{\top}\boldsymbol{\beta}_{0}\boldsymbol{)+}%
s^{\prime}\boldsymbol{(}\mathbf{x}_{i}^{\top}\boldsymbol{\beta}_{0}%
\boldsymbol{)\mathbf{x}_{i}^{\top}(\beta-\beta}_{0}). \label{APS}%
\end{equation}
Then an approximate value to the LST estimator can be found by the value
$\boldsymbol{\beta}_{1}$ that minimizes
\[
\sum_{i=1}^{n}\left(  t(y_{i})-s\boldsymbol{(}\mathbf{x}_{i}^{\top
}\boldsymbol{\beta}_{0}\boldsymbol{)-}s^{\prime}\boldsymbol{(}\mathbf{x}%
_{i}^{\top}\boldsymbol{\beta}_{0}\boldsymbol{)\mathbf{x}_{i}^{\top}%
(\beta-\beta}_{0})\right)  ^{2}\ \ .
\]
Therefore $\boldsymbol{\beta}_{1}-\boldsymbol{\beta}_{0}$ is the LS estimator for
a linear model with responses $t(y_{1}),\ldots,t(y_{n})$ \ and regressor vectors
$s^{\prime}\boldsymbol{(}\mathbf{x}_{1}^{\top}\boldsymbol{\beta}%
_{0}\boldsymbol{)\mathbf{x}_{1}},\ldots,s^{\prime}\boldsymbol{(}\mathbf{x}%
_{n}^{\top}\boldsymbol{\beta}_{0}\boldsymbol{)\mathbf{x}_{n}\ }$and
consequently
\begin{equation}
\boldsymbol{\beta}_{1}=\boldsymbol{\beta}_{0}+\left(  \mathbf{X}^{\top
}\mathbf{W(\boldsymbol{X}}\boldsymbol{\beta}\mathbf{_{0})}^{2}\mathbf{X}%
\right)  ^{-1}\mathbf{\ \mathbf{X}^{\top}W\mathbf{(X}}\boldsymbol{\beta}%
_{0}\mathbf{)(T-s(X}\boldsymbol{\beta}_{0}\mathbf{)),} \label{ALS}%
\end{equation}
where $\mathbf{X}$ is the $n\times p$ matrix whose $i$-th row is
$\boldsymbol{\mathbf{x}_{i}^{\top},}$  $\mathbf{s(X}\boldsymbol{\beta
}\mathbf{)}=(s(\mathbf{x}_{1}^{\top}\boldsymbol{\beta}),\ldots,s(\mathbf{x}%
_{n}^{\top}\boldsymbol{\beta}))^{\top}$, $\mathbf{W\mathbf{(X}}\boldsymbol{\beta
}\mathbf{)}$ is the diagonal matrix with diagonal elements $s^{\prime
}(\mathbf{x}_{1}^{\top}\boldsymbol{\beta}),\ldots,s^{\prime}(\mathbf{x}_{n}%
^{\top}\boldsymbol{\beta})$ and $\mathbf{T=(}t(y_{1}),\ldots,t(y_{n}))^{\top}.$

An iterative procedure to compute the LST estimator can be obtained putting
\begin{equation}
\boldsymbol{\beta}_{k+1}=\boldsymbol{\beta}_{k}+\left(  \mathbf{X}^{\top
}\mathbf{W(\boldsymbol{X}}\boldsymbol{\beta}_{k}\mathbf{)}^{2}\mathbf{X}%
\right)  ^{-1}\mathbf{\ \mathbf{X}^{\top}W\mathbf{(X}}\boldsymbol{\beta}%
_{k}\mathbf{)(T-s(X}\boldsymbol{\beta}_{k}\mathbf{)),} \label{IALS}%
\end{equation}
and stopping when $\left\Vert \boldsymbol{\beta}_{k+1}-\boldsymbol{\beta
}_{k+1}\right\Vert /\left\Vert \boldsymbol{\beta}_{k}\right\Vert \leq\delta,$
where $\delta$ is the error tolerance.

Suppose that $\boldsymbol{\beta}_{k}$ converges to $\boldsymbol{\beta}^{\ast},$
then this value should satisfy the \ LST estimating equation . In fact, taking
limit in both sides of (\ref{IALS}) we get
\[
\boldsymbol{\ }\left(  \mathbf{X}^{\top}\mathbf{W(\boldsymbol{X}%
}\boldsymbol{\beta}^{\ast}\mathbf{)}^{2}\mathbf{X}\right)  ^{-1}%
\mathbf{\mathbf{X}^{\top}W\mathbf{(X}}\boldsymbol{\beta}^{\ast}\mathbf{)(T-s(}%
\boldsymbol{\beta}^{\ast}\mathbf{))=0,}%
\]
which is equivalent to
\[
\mathbf{\mathbf{X}^{\top}W\mathbf{(X}}\boldsymbol{\beta}^{\ast}\mathbf{)}%
^{\top}\mathbf{(T-s(}\boldsymbol{\beta}^{\ast}\mathbf{))=0,}%
\]
and then $\boldsymbol{\beta}^{\ast}$ satisfies  the estimating equation of the
LST estimator.

To start the algorithm, it will be convenient to write equation (\ref{ALS}) in
the following slightly different way%
\begin{equation}
\boldsymbol{\beta}_{1}=\left(  \mathbf{X}^{\top}\mathbf{W(\boldsymbol{X}%
}\boldsymbol{\beta}\mathbf{_{0})}^{2}\mathbf{X}\right)  ^{-1}(\mathbf{X}%
^{\top}\mathbf{W(\boldsymbol{X\beta}_{0})}^{2}\mathbf{X}\boldsymbol{\beta}%
_{0}+\ \mathbf{\mathbf{X}^{\top}W\mathbf{(X}}\boldsymbol{\beta}_{0}%
\mathbf{)(T-s(X}\boldsymbol{\beta}_{0}\mathbf{)).}\label{ALS1}%
\end{equation}

Observe that according to (\ref{ALS1}) to compute $\boldsymbol{\beta}_{1}$
\ we only need to give $\boldsymbol{\eta}_{0}=\mathbf{\boldsymbol{X}%
}\boldsymbol{\beta}_{0}$.\ Then, $\ $since for Poisson regression and log link
it holds  $\boldsymbol{\mathbf{x}_{i}^{\top}\beta}$ $=\log(E(y_{i})),$ it
seems reasonable to take $\boldsymbol{\eta}_{0}=(\log(y_{1}+0.1),\ldots,\log
(y_{n}+0.1))^{\top}.$ The value 0.1 is added to avoid numerical problem when
$y_{i}=0.$ To compute the estimators $\boldsymbol{\hat{\beta}}_{(j)}$ \ only
one iteration is performed. The reason is \ that for these auxiliary
estimators the accuracy is not as important as the speed at which they can be
computed. \ Our experiments show that there is no noticeable loss in the
precision of the final estimate by doing this but, on the other hand, the
computation times decrease significantly. \ \ \ \ 

We describe now an analogous iterative algorithm for computing the MT
estimator. Suppose that we have an initial robust estimator $\boldsymbol{\beta
}_{0}.$ We compute a new value using two approximations. As in the case of the
LST estimator, replacing, in \eqref{Mest1}, $m\left(  g^{-1}\left(
\mathbf{x}_{i}^{\top}\boldsymbol{\beta}\right)  \right)  $ by (\ref{APS}) we
consider the approximate loss function
\[
\sum_{i=1}^{n}\rho\left(  t(y_{i})-s\boldsymbol{(}\mathbf{x}_{i}^{\top
}\boldsymbol{\beta}_{0}\boldsymbol{)-}s^{\prime}\boldsymbol{(}\mathbf{x}%
_{i}^{\top}\boldsymbol{\beta}_{0}\boldsymbol{)\mathbf{x}_{i}^{\top}%
(\beta-\beta}_{0})\right)  \ \ .
\]

Differentiating with respect to $\boldsymbol{\beta}$ $\ $we obtain the
estimating equation
\begin{equation}
\sum_{i=1}^{n}\psi\left(  t(y_{i})-s\boldsymbol{(}\mathbf{x}_{i}^{\top
}\boldsymbol{\beta}_{0}\boldsymbol{)-}s^{\prime}\boldsymbol{(}\mathbf{x}%
_{i}^{\top}\boldsymbol{\beta}_{0}\boldsymbol{)\mathbf{x}_{i}^{\top}%
(\beta-\beta}_{0})\right)  s^{\prime}\boldsymbol{(}\mathbf{x}_{i}^{\top
}\boldsymbol{\beta}_{0}\boldsymbol{)\mathbf{x}_{i}=0}, \label{MTE}%
\end{equation}
where $\psi=\rho^{\prime}.$ Note that this equation can be written as%
\begin{equation}
\sum_{i=1}^{n}\ \left(  t(y_{i})-s\boldsymbol{(}\mathbf{x}_{i}^{\top
}\boldsymbol{\beta}_{0}\boldsymbol{)-}s^{\prime}\boldsymbol{(}\mathbf{x}%
_{i}^{\top}\boldsymbol{\beta}_{0}\boldsymbol{)\mathbf{x}_{i}^{\top}%
(\beta-\beta}_{0})\right)  w(\mathbf{x}_{i}^{\top}\boldsymbol{\beta
},\mathbf{x}_{i}^{\top}\boldsymbol{\beta}_{0})s^{\prime}\boldsymbol{(}%
\mathbf{x}_{i}^{\top}\boldsymbol{\beta}_{0}\boldsymbol{)\mathbf{x}_{i}},
\label{MTA1}%
\end{equation}
where
\[
w(u,v)=\frac{\psi\left(  t(y_{i})-s\boldsymbol{(}v\boldsymbol{)-}s^{\prime
}\boldsymbol{(}v\boldsymbol{)}\left(  u-v\right)  \right)  }{t(y_{i}%
)-s\boldsymbol{(}v\boldsymbol{)-}s^{\prime}\boldsymbol{(}v\boldsymbol{)}%
\left(  u-v\right)  }.
\]
Since $\boldsymbol{\beta}$ \ should be close to $\boldsymbol{\beta}_{0},$ the
second approximation is to replace, in (\ref{MTA1}), $w(\mathbf{x}_{i}^{\top
}\boldsymbol{\beta},\mathbf{x}_{i}^{\top}\boldsymbol{\beta}_{0})$ by $w^{\ast
}(\mathbf{x}_{i}^{\top}\boldsymbol{\beta}_{0})=w(\mathbf{x}_{i}^{\top
}\boldsymbol{\beta}_{0},\mathbf{x}_{i}^{\top}\boldsymbol{\beta}_{0}).$ Then
$\boldsymbol{\beta}_{1}$ is defined as the solution of the approximate
estimating equation
\[
\sum_{i=1}^{n}\ \left(  t(y_{i})-s\boldsymbol{(}\mathbf{x}_{i}^{\top
}\boldsymbol{\beta}_{0}\boldsymbol{)-}s^{\prime}\boldsymbol{(}\mathbf{x}%
_{i}^{\top}\boldsymbol{\beta}_{0}\boldsymbol{)\mathbf{x}_{i}^{\top}%
(\beta-\beta}_{0})\right)  w^{\ast}(\mathbf{x}_{i}^{\top}\boldsymbol{\beta
}_{0})s^{\prime}\boldsymbol{(}\mathbf{x}_{i}^{\top}\boldsymbol{\beta}%
_{0}\boldsymbol{)\mathbf{x}_{i}}%
\]
and is given by%
\[
\boldsymbol{\beta}_{1}=\boldsymbol{\beta}_{0}+\left(  \mathbf{X}^{\top
}\mathbf{W}^{2}\mathbf{(\boldsymbol{X}}\boldsymbol{\beta}_{0}\mathbf{)^{\top
}\mathbf{W}}^{\ast}\mathbf{\mathbf{(\boldsymbol{X}}}\boldsymbol{\beta}%
_{0}\mathbf{\mathbf{)\ }X}\right)  ^{-1}\mathbf{\ \mathbf{X}^{\top
}W\mathbf{(X}}\boldsymbol{\beta}_{0}\mathbf{)\mathbf{\mathbf{W}}^{\ast
}\mathbf{\mathbf{(\boldsymbol{X}}}}\boldsymbol{\beta}_{0}%
\mathbf{\mathbf{\mathbf{)}}(T-s(X}\boldsymbol{\beta}_{0}\mathbf{)),}%
\]
where $\mathbf{W}^{\ast}\mathbf{(X}\boldsymbol{\beta}\mathbf{)}$ is the
$n\times n$ diagonal matrix with diagonal elements $w^{\ast}%
(\boldsymbol{\mathbf{x}}_{1}^{\top}\boldsymbol{\beta}),\ldots,w^{\ast
}(\boldsymbol{\mathbf{x}}_{n}^{\top}\boldsymbol{\beta}).$

Then, the iterative procedure to compute the MT estimator is given by
\begin{equation}
\boldsymbol{\beta}_{k+1}=\boldsymbol{\beta}_{k}+\left(  \mathbf{X}^{\top
}\mathbf{W}^{2}\mathbf{(\boldsymbol{X}}\boldsymbol{\beta}_{k}\mathbf{)^{\top
}\mathbf{W}}^{\ast}\mathbf{\mathbf{(\boldsymbol{X}}}\boldsymbol{\beta}%
_{k}\mathbf{\mathbf{)\ }X}\right)  ^{-1}\mathbf{\ \mathbf{X}^{\top
}W\mathbf{(X}}\boldsymbol{\beta}_{k}\mathbf{)\mathbf{\mathbf{W}}^{\ast
}\mathbf{\mathbf{(\boldsymbol{X}}}}\boldsymbol{\beta}_{k}%
\mathbf{\mathbf{\mathbf{)}}(T-s(X}\boldsymbol{\beta}_{k}\mathbf{)),}
\label{IMT}%
\end{equation}

Suppose that $\boldsymbol{\beta}_{k}\rightarrow\boldsymbol{\beta}^{\ast},$ then
taking limits in both sides of (\ref{IMT}), we get%
\[
\boldsymbol{\ }\mathbf{\ \mathbf{X}^{\top}W\mathbf{(X}}\boldsymbol{\beta
}^{\ast}\mathbf{)\mathbf{\mathbf{W}}^{\ast}\mathbf{\mathbf{(\boldsymbol{X}}}%
}\boldsymbol{\beta}^{\ast}\mathbf{\mathbf{\mathbf{)}}(T-s(X}\boldsymbol{\beta
}^{\ast}\mathbf{))}=\mathbf{0,}%
\]
and this is equivalent to
\begin{equation}
\mathbf{\mathbf{X}^{\top}W\mathbf{(X}}\boldsymbol{\beta}^{\ast}\mathbf{)\Psi
}^{\ }\mathbf{(\mathbf{\mathbf{\boldsymbol{X}}}}\boldsymbol{\beta}^{\ast
}\mathbf{)}=\mathbf{0,} \label{MMTE}%
\end{equation}
where $\mathbf{\Psi}^{\ }\mathbf{(\mathbf{\mathbf{\boldsymbol{X}}}%
}\boldsymbol{\beta}\mathbf{)=(}\psi(t(y_{1})-s(\mathbf{x}_{i}^{\top
}\boldsymbol{\beta})),\ldots,\psi(t(y_{n})-s(\mathbf{x}_{i}^{\top}%
\boldsymbol{\beta}))^{\top}$. Then $\boldsymbol{\beta}^{\ast}$ satisfies the
estimating equation of the MT estimator.


\begin{thebibliography}{10}
\providecommand{\natexlab}[1]{#1}
\providecommand{\url}[1]{\texttt{#1}}
\expandafter\ifx\csname urlstyle\endcsname\relax
  \providecommand{\doi}[1]{doi: #1}\else
  \providecommand{\doi}{doi: \begingroup \urlstyle{rm}\Url}\fi

\bibitem[Alqallaf and Agostinelli(2016)]{AlqallafAgostinelli2016}
F.~Alqallaf and C.~Agostinelli.
\newblock Robust inference in generalized linear models.
\newblock \emph{Communications in Statistics - Simulation and Computation},
  45\penalty0 (9):\penalty0 3053--3073, 2016.
\newblock \doi{10.1080/03610918.2014.911896}.

\bibitem[Bergesio and Yohai(2011)]{BergesioYohai2011}
A.~Bergesio and V.J. Yohai.
\newblock Projection estimators for generalized linear models.
\newblock \emph{Journal of the American Statistical Association}, 106:\penalty0
  661--671, 2011.

\bibitem[Bianco et~al.(2013)Bianco, Boente, and Rodrigues]{Biancoetal2013}
A.M. Bianco, G.~Boente, and I.M. Rodrigues.
\newblock Resistant estimators in poisson and gamma models with missing
  responses and an application to outlier detection.
\newblock \emph{Journal of Multivariate Analysis}, 114:\penalty0 209--226,
  2013.

\bibitem[Cantoni and Ronchetti(2001)]{CantoniRonchetti2001}
E.~Cantoni and E.~Ronchetti.
\newblock Robust inference for generalized linear models.
\newblock \emph{Journal of the American Statistical Association}, 96:\penalty0
  1022--1030, 2001.

\bibitem[Cook(1977)]{Cook1977}
R.D. Cook.
\newblock Detection of influential observation in linear regression.
\newblock \emph{Technometrics}, 19\penalty0 (1):\penalty0 15--18, 1977.
\newblock \doi{10.2307/1268249}.

\bibitem[K\"{u}nsch et~al.(1989)K\"{u}nsch, Stefanski, and
  Carroll]{Kunschetal1989}
H.~K\"{u}nsch, L.~Stefanski, and R.~Carroll.
\newblock Conditionally unbiased bounded-influence estimation in general
  regression models, with applications to generalized linear models.
\newblock \emph{Journal of the American Statistical Association}, 84:\penalty0
  460--466, 1989.

\bibitem[McCullagh and Nelder(1989)]{McCullaghNelder1989}
P.~McCullagh and J.A. Nelder.
\newblock \emph{Generalized Linear Models}.
\newblock Chapman and Hall/CRC, second edition, 1989.

\bibitem[Pe{\~n}a and Yohai(1999)]{PenaYohai1999}
D.~Pe{\~n}a and V.J. Yohai.
\newblock A fast procedure for outlier diagnostics in large regression
  problems.
\newblock \emph{Journal of the American Statistical Association}, 94:\penalty0
  434--445, 1999.

\bibitem[Rousseeuw and Leroy(1987)]{RousseeuwLeroy1987}
P.J. Rousseeuw and A.M. Leroy.
\newblock \emph{Robust regression and outlier detection}.
\newblock Wiley and Sons, 1987.

\bibitem[Valdora and Yohai(2014)]{ValdoraYohai2014}
M.~Valdora and V.J. Yohai.
\newblock Robust estimators for generalized linear models.
\newblock \emph{Journal of Statistical Planning and Inference}, 146:\penalty0
  31--48, 2014.

\end{thebibliography}
\end{document}